\documentclass[aps,pra,twocolumn,superscriptaddress,showpacs
]{revtex4}

\usepackage[english]{babel}
\usepackage{graphicx}
\usepackage{amsmath}
\usepackage{amssymb}
\usepackage{color}

\usepackage{cancel}

\begin{document}

\title{Generalized minimal output entropy conjecture for  one-mode Gaussian channels : \\ definitions and  some exact results}

\author{V. Giovannetti}
\affiliation {NEST,  Scuola Normale Superiore \& CNR-INFM, Piazza   dei Cavalieri 7, I-56126 Pisa, Italy}
\author{A. S. Holevo}
\affiliation{Steklov Mathematical Institute, Gubkina 8, 119991 Moscow, Russia}
\author{S. Lloyd} \affiliation{Research Laboratory of Electronics, Massachusetts Institute of Technology, Cambridge, MA, 02139, USA}
\author {L. Maccone}
\affiliation{Research Laboratory of Electronics, Massachusetts Institute of Technology, Cambridge, MA, 02139, USA}

\date{\today}

\begin{abstract}
A formulation of the generalized minimal output entropy conjecture
for Gaussian channels is presented. It asserts that, for states with
fixed input entropy, the minimal value of the output entropy of the
channel (i.e.  the minimal output entropy increment for fixed input
entropy)  is achieved by Gaussian states.
In the case of centered channels (i.e. channels which do not add
squeezing to the input state) this implies that the minimum is
obtained by thermal (Gibbs) inputs. The conjecture is proved to be
valid in some special cases.

\end{abstract}


\maketitle

\section{Introduction}

Several of the most difficult problems in quantum information
theory~\cite{BENSHOR,PETZ} deal with optimizations of nonlinear cost
functions. In particular, in close analogy to what is done in the
classical theory~\cite{COVER}, the efficiency of a communication
line (quantum channel) is measured by  maximizing an entropic
functional  over the set of possible channel inputs. Apart from some
special cases, such optimizations are in general too complex to be
performed explicitly.  In an effort to simplify the analysis,
several conjectures were proposed based either on physical intuition
or on symmetries of the problem. The most known is the additivity
conjecture recently disproved by Hastings~\cite{HASTINGS}: it
claimed that the minimal value of the von Neumann entropy at the
output of a memoryless channel is achieved by separable input states
or, equivalently~\cite{SHOR}, that its Holevo capacity~\cite{HOLEVO}
is additive.  As a consequence we now know  that  the classical
capacity~\cite{HWS}  of a memoryless quantum channel (i.e. the
maximum achievable rate of reliable classical communication) will be
in general difficult to evaluate as it necessarily requires to
perform a nontrivial regularization over infinitely many channel
uses.

Of course it is still possible that  special classes of quantum
channels will obey additivity rules that would allow us to
simplify their analysis. In particular it is generally believed that
Bosonic Gaussian channels~\cite{HW,CEGH,EW} should be one of  such
classes. As a matter of fact Bosonic Gaussian channels  appear to
have a preferred (simple) set of inputs states  (the Gaussian
states~\cite{BOOK}) over which the optimization of the relevant
entropic quantities can be performed simplifying the
calculation~\cite{EXT}. Several results support such belief. In
particular the capacity of lossy Gaussian channels was proved to be
additive by  explicitly computing its value~\cite{CAP}; for
thermalizing channels the minimum values of the  R\'{e}nyi entropies
were shown to be additive for integer orders and unconstrained
input~\cite{REN}, and for arbitrary order under the constraint of
Gaussian inputs~\cite{REN1}; finally the degradability and
additivity of the coherent information for some of those channels
was established in Refs.~\cite{CANONICAL1,CANONICAL,WOLF1}.
Partially motivated by the above results few years ago a conjecture
was proposed  a solution of which would allow one to simplify the
whole scenario, allowing for instance a direct computation of the
classical capacity of some Gaussian channels. In particular it was
suggested that the (unconstrained) minimum of the von
Neumann~\cite{CONJEC} or R\'{e}nyi~\cite{CONJEC1} entropies for
attenuators or additive Gaussian classical noise channels should be
achieved by the vacuum state. 
 Up to now all attempts to prove this
apparently innocuous claim have failed, including an innovative
approach that was recently presented~\cite{PAPER} in which
the original conjecture of Ref.~\cite{CONJEC}  was generalized. 
 The aim of this paper is to review the problem 
extending the   conjecture  to include
all Gaussian channels, and to prove it  in some special but not
necessarily trivial cases.
In our treatment of unbounded operators arising from Canonical
Commutation Relations we focus on the aspects essential for physical
calculations. A number of analytical complications related to
infinite dimensionality and unboundedness unavoidably arises in
connection with Bosonic systems and Gaussian states. A detailed
treatment of related mathematical tools can be found in \cite{ho1,wehrl}
and the references therein.

The paper is organized as follows. In Sec.~\ref{s:conj} the
notations are introduced  and the  formulation of the conjecture is
presented. For channels which admit semigroup structure we also introduce an  infinitesimal version of the conjecture. 
In Sec.~\ref{s:special} we focus on a special class of channels and
introduce a relatively simple argument based on subadditivity
properties of the von Neumann entropy which allows one to prove the
conjecture for some (lucky) cases. Sec.~\ref{singular}  proves the
conjecture for a class of degenerate Gaussian channels. Conclusion
and remarks are given in Sec.~\ref{s:conc}.

\section{The conjecture} \label{s:conj}
Let  $\Phi$ be a linear,  completely positive, trace preserving
(LCPT)  map (see, e.g.~\cite{PETZ})  which transforms  a (possibly
infinite dimensional) input system $A$ to an output system $B$.
Indicating with $\mathfrak{S} ({\cal H}_A)$ the set of density
matrices of the input space, we define $\mathfrak{S}_{S_0}({\cal
H}_A)$  and $\mathfrak{S}^+_{S_0}({\cal H}_A)$ as the subsets of
$\mathfrak{S}({\cal H}_A)$ formed, respectively, by states with
entropy {\em equal} to $S_0$ and by states with entropy {\em larger
than or equal to} $S_0$, i.e.
\begin{eqnarray}
\mathfrak{S}_{S_0}({\cal H}_A) =\{ \rho \in \mathfrak{S}({\cal H}_A) : S(\rho) =S_0\}\;, \\
\mathfrak{S}^+_{S_0}({\cal H}_A) =\{ \rho \in \mathfrak{S}({\cal H}_A) : S(\rho) \geqslant S_0\}\;,
\end{eqnarray}
with  $S(\rho) = -\mbox{Tr} [ \rho \ln \rho] $ being the von Neumann
entropy~\cite{PETZ} of $\rho$. By the concavity of $S$ one has that
the $\mathfrak{S}^+_{S_0}({\cal H}_A)$ are convex sets which can be
expressed as proper unions of the $\mathfrak{S}_{S_0}({\cal H}_A)$,
namely
 $\mathfrak{S}^+_{S_0}({\cal H}_A) =   \bigcup_{S\geqslant S_0} \mathfrak{S}_{S}({\cal H}_A)$. Furthermore they  form an ordered family under inclusion, i.e.
\begin{eqnarray}
 \mathfrak{S}^+_{S_0}({\cal H}_A) \subset  \mathfrak{S}^+_{S_0'}({\cal H}_A)\;, \qquad  \mbox{for all $S_0' >  S_0$}.
 \end{eqnarray}
 In particular for $S_0=0$, $\mathfrak{S}_{0}({\cal H}_A)$ represents  the set of pure state of the system while $\mathfrak{S}^+_{0}({\cal H}_A)$ coincides with
 the whole space  $\mathfrak{S} ({\cal H}_A)$.
We are interested in computing the minimum value that the  output entropy $S(\Phi(\rho))$ can take on the set $\mathfrak{S}_{S_0}({\cal H}_A)$, i.e. the quantity
\begin{eqnarray} \label{defiF}
{\cal F}(\Phi; S_0) \equiv \inf_{\rho \in \mathfrak{S}_{S_0}({\cal H}_A)} S(\Phi(\rho))\;.
\end{eqnarray}
Due to the concavity of $S$ and the linearity of $\Phi$, such a minimum can also be expressed as a minimum over the larger set $\mathfrak{S}^+_{S_0}({\cal H}_A)$, i.e.
\begin{eqnarray} \label{defiF1}
{\cal F}(\Phi; S_0) = \inf_{\rho \in \mathfrak{S}^+_{S_0}({\cal H}_A)} S(\Phi(\rho))\;.
\end{eqnarray}
For $S_0=0$ the quantity~(\ref{defiF}) provides the (unconstrained)
minimal output entropy of the channel which plays a fundamental role
in quantum communication~\cite{BENSHOR}. (In particular, its
additivity property under successive uses of the channel was
recently disproved in Ref.~\cite{HASTINGS}.) Moreover, in the
special case in which $\Phi$ represents an attenuator or additive
Gaussian classical noise channels~\cite{HW} operating on a single
Bosonic mode, a conjecture was proposed in Ref.~\cite{CONJEC} which,
if true, would allow one to compute in closed form its classical
capacity~\cite{HWS} under the energy constraint. Specifically it was
conjectured that the  value of ${\cal F}(\Phi; S_0=0)$  is attained
by a Gaussian input state. In a recent attempt~\cite{PAPER}  to
prove such a property, it was recently extended  to include  all
values of $S_0 >0$ and a broader class of maps.
 Indeed consider a set of $n$ input Bosonic modes and a Gaussian channel $\Phi$ which maps them into $m$ output modes. We remind that
for these systems a state is said to be Gaussian if 
its symmetrically ordered characteristic function (or equivalently its Wigner distribution) corresponds to a  Gaussian envelop~\cite{BOOK}, whereas
 a LCPT map is said to be Gaussian channel if, when 
 acting on a Gaussian state of the input modes transforms it into an Gaussian state of the output modes~\cite{HW,CEGH,EW,BOOK}.
It is claimed that:  \\
\newline
{\bf Conjecture (v1):} {\em For all $S_0 \geqslant 0$ the minimization in {\em Eq.~(\ref{defiF})} is saturated by
a Gaussian element of the set {\em $\mathfrak{S}_{S_0}({\cal H}_A)$}, i.e. }
\begin{eqnarray}
 {\cal F}(\Phi; S_0)=  S(\Phi(\rho_0))\;, \label{CONJ}
\end{eqnarray}
{\em with $\rho_0\in \mathfrak{S}_{S_0}({\cal H}_A)$ a Gaussian
state (notice that for all  $S_0$, the sets
$\mathfrak{S}_{S_0}({\cal H}_A)$ always admit at least one Gaussian
element).}
\newline

While in some simple cases the conjecture can be easily verified, in
the general scenario it appears to be particularly challenging. In
the following we will specify the analysis to the case of
single-mode Gaussian channels ($n=m=1$) for which the canonical
decomposition of $\Phi$ applies~\cite{CANONICAL,CANONICAL1}. In
particular, it is known that apart from the special cases which we
will treat  in Sec.~\ref{singular}, by making a proper choice
of the canonical observables at the input and the output of the
channel one can focus on   {\em centered} Gaussian channels which
respect the standard complex structure associated with the
multiplication by $i$ (these channels do not introduce squeezing or
displacement). 
They have the property to induce the following transformation on the average  photon expectation value, 
\begin{eqnarray}\label{dfe}
\mbox{Tr} [ \Phi(\rho) a^\dag a ] =  \kappa^2 \; \mbox{Tr}[ \rho a^\dag a]  +  c \;,
\end{eqnarray}
where  $\kappa$ and $c$ are constants which depend upon $\Phi$,
where $a, a^\dag$ are the annihilation and creation operator of the
system mode.
Specifically  attenuator channels are characterized by  $\kappa^2 \in [ 0,1]$
and $c = (1-\kappa^2) N$ with $N \geqslant 0$,  
while amplifier channels are characterized by 
 $\kappa^2
\geqslant 1$ and $c= (\kappa^2 -1)(N+1)$ where  again $N \geqslant 0$,  (class $C$
of~\cite{CANONICAL}).
For additive Gaussian classical noise channel
(class $B_2$ of~\cite{CANONICAL}) instead one has $\kappa=1$ and
$c=N$. Finally for the weak conjugate of the amplifier channels
(class $D$ of~\cite{CANONICAL})  one has $\kappa^2 \geqslant 0$ and
$c= \kappa^2(N+1)+N$. Due to the property (\ref{dfe}) one can
refine the conjecture {\bf (v1)} by saying that the state $\rho_0\in
\mathfrak{S}_{S_0}({\cal H})$ entering in Eq.~(\ref{CONJ}) is the
(thermal) Gibbs state,
\begin{eqnarray} \label{FFD}
\rho_0 = \frac{1}{N_0+1} \; \left(\frac{N_0}{N_0 +1}\right)^{a^\dag a} \;,
\end{eqnarray}
with $N_0= \mbox{Tr} [ a^\dag a \rho_0] > 0$ being the average photon
number of $\rho_0$ which allows to express the input entropy of
$\rho_0$ as
\begin{eqnarray}\label{entropyN}
S_0  = g(N_0) \equiv (N_0+1) \ln (N_0+1) - N_0 \ln N_0,
\end{eqnarray}
   (for $N_0=0$ the the density matrix  $\rho_0$ must be identified with the vacuum state, while $S_0=0$).
By general properties of Gibbs states, see e.g. \cite{wehrl},
we know that $\rho_0$ is the element of $\mathfrak{S}^+_{S_0}({\cal
H})$ which has minimal energy, i.e.
\begin{eqnarray}
N_0 \leqslant \mbox{Tr} [ a^\dag a \rho]  \;,\qquad\qquad \mbox{for all $\rho\in \mathfrak{S}^+_{S_0}({\cal H})$},
\end{eqnarray}
the identity applying only for $\rho=\rho_0$. Furthermore, for the channel under consideration
$\Phi$ will transform $\rho_0$ into a new Gibbs state $\rho_0'=\Phi(\rho_0)$, having mean photon number $N'_0= \kappa^2 N_0 + c$ with $\kappa,c$ as in Eq.~(\ref{dfe}). Therefore the
conjecture can be restated as follows:
\newline \\
{\bf Conjecture (v2):} {\em For all $S_0 \geqslant 0$ the minimization in {\em Eq.~(\ref{defiF})} is saturated by the Gibbs state $\rho_0$ of {\em Eq.~(\ref{FFD})}.Therefore
it holds}
\begin{eqnarray}
 {\cal F}(\Phi; S_0)  = S(\Phi(\rho_0))= g(\kappa^2 N_0 + c)\;,  \label{CONJ1}
\end{eqnarray}
{\em or, equivalently, for all $\rho\in \mathfrak{S}_{S_0}({\cal
H})$ one has}
\begin{eqnarray}
S(\Phi(\rho))\geqslant  g(\kappa^2 N_0 + c)\;.
\label{CONJ2}
\end{eqnarray}
The inequality~(\ref{CONJ2}) can be cast in a different form by
introducing the relative entropy $S(\rho||\sigma)$, see e.g.
\cite{wehrl}. Indeed, for all $\rho \in
\mathfrak{S}_{S_0}({\cal H})$ simple algebraic manipulations  allows
us to write
\begin{eqnarray}
&&S(\Phi(\rho)) - g(\kappa^2 N_0 + c) \label{EQUIV}\\
&&\qquad = \kappa^2\; \frac{\ln\left(\tfrac{ \kappa^2 N_0 + c +1}{\kappa^2 N_0 + c}\right)}{ \ln\left(\tfrac{ N_0 + 1}{ N_0 }\right)} S(\rho||\rho_0) - S(\Phi(\rho)||\Phi(\rho_0))\;,
 \nonumber
\end{eqnarray}
where we used the fact that $S(\rho)=S_0$. This shows that  a necessary and sufficient condition for the  conjecture~(\ref{CONJ2}) is the inequality,
\begin{eqnarray}
\kappa^2 \frac{\ln\left(\tfrac{ \kappa^2 N_0 + c +1}{\kappa^2 N_0 + c}\right)}{ \ln\left(\tfrac{ N_0 + 1}{ N_0 }\right)} S(\rho||\rho_0) \geqslant  S(\Phi(\rho)||\Phi(\rho_0))\;, \label{dds}
\end{eqnarray}
which needs to apply to all input states $\rho\in
\mathfrak{S}_{S_0}({\cal H})$. It is worth reminding that the
relative entropy is monotonically decreasing under the action of
LCPT maps (see, e.g.~\cite{PETZ}), that is $S(\rho||\rho_0)
\geqslant S(\Phi(\rho)||\Phi(\rho_0))$ for all $\Phi$, $\rho$ and
$\rho_0$. Therefore a sufficient condition to prove  Eq.~(\ref{dds})
would be $\kappa^2 {\ln\left(\tfrac{ \kappa^2 N_0 + c +1}{\kappa^2
N_0 + c}\right)}\geqslant { \ln\left(\tfrac{ N_0 + 1}{ N_0
}\right)}$. Unfortunately however, for all values of $\kappa$, $N_0$
and $c$
as in Eq.~(\ref{dfe})  this inequality is always false 
(notice that $c$ and $k$ cannot be taken as independent variables).
Incidentally this shows that proving Eq.~(\ref{dds}) (and thus the
conjecture) requires one to go beyond the monotonicity property of
the relative entropy.

For those Gaussian channels $\Phi$ which  possess  a semigroup
structure~\cite{SEMIGROUP} the conjecture can be
rephrased in terms of a condition on the infinitesimal increments of
the entropy. Specifically let ${\cal L}$ be a Lindblad generator and
let $\{ \Phi_t: t\geqslant 0\}$ be  a one-parameter family of
Gaussian LCPT maps which solve the equation
\begin{eqnarray}\label{Semi}
\frac{ \partial}{\partial t}\Phi_t  = {\cal L} \circ {\Phi_t} \;,  \qquad \qquad
\Phi_0 = {\cal I} \;,
\end{eqnarray}
with $\cal I$ being the identity channel and $\circ$ being the
composition of maps. For instance this property holds for
attenuator, amplifier and additive Gaussian classical noise channels
with
\begin{eqnarray}\label{ffd111}
{\cal L} = \frac{\gamma_+}{2} \;{\cal L}_+ + \frac{\gamma_-}{2} \; {\cal L}_-\;,
\end{eqnarray}
where $\gamma_\pm$ are positive parameters related to $\kappa$ and $c$ of Eq.~(\ref{dfe}) and where
\begin{eqnarray}
{\cal L}_+(\cdot)= 2 a^\dag (\cdot) a - a a^\dag (\cdot) - (\cdot) a a^\dag \;,\\
{\cal L}_-(\cdot)= 2 a (\cdot) a^\dag - a^\dag a(\cdot) - (\cdot) a^\dag a \;.
\end{eqnarray}
In particular one can easily verify that   attenuators form a semigroup satisfying Eq.~(\ref{Semi}) with $\gamma_+= N$, $\gamma_-= N+1$ when setting $\kappa^2= e^{-t}$.
Similarly for the amplifiers we have      $\gamma_+= N+1$, $\gamma_-= N$ by choosing $\kappa^2= e^{t}$, while
for additive Gaussian classical noise channels we have $\gamma_+=\gamma_-=1$ when setting $N=t$.

For each $T\geqslant0$ we can
then express the output entropy of the map $\Phi_{t=T}$ in the  integral form
\begin{eqnarray} \label{diff3}
S(\Phi_{T}(\rho)) = S(\rho) - \int_0^T dt \;\;\mbox{Tr}[ {\cal L} (\Phi_t (\rho)) \ln \Phi_t (\rho) ]\;,
\end{eqnarray}
where we used the fact that $\tfrac{\partial}{\partial t}
S(\Phi_{t}(\rho)) = -\mbox{Tr}[ {\cal L} (\Phi_t (\rho)) \ln \Phi_t
(\rho) ]$. Due to the identity~(\ref{diff3}) we can now  present an
infinitesimal version of the conjecture:
\newline \\
{\bf Conjecture (v3):} {\em For all $S_0 \geqslant 0$ define the quantity}
\begin{eqnarray}
{\cal F}({\cal L}; S_0)  \equiv - \inf_{\rho\in \mathfrak{S}_{S_0}({\cal H}_A)} \mbox{Tr}[ {\cal L} (\rho) \ln \rho  ] \;,
\end{eqnarray}
{\em  then the minimization is saturated by the Gibbs state $\rho_0\in  \mathfrak{S}_{S_0}({\cal H}_A)$ defined in {\em Eq.~(\ref{FFD})}, i.e. it holds the identity}
\begin{eqnarray}
{\cal F}({\cal L}; S_0) &=& - \mbox{Tr}[ {\cal L} (\rho_0) \ln \rho_0  ] \nonumber \\
&=& [(\gamma_+-\gamma_-) N_0 + \gamma_+]  \; \ln(\tfrac{N_0+1}{N_0})
\;.\label{infinitesimal}
\end{eqnarray}

Clearly if the conjecture {\bf (v2)} is true for all discrete  channels $\Phi_T$ and for  all $S_0$,
then also the infinitesimal version {\bf (v3)} of the conjecture must be true.
This follows simply by the fact that
the functional ${\cal F}({\cal L}; S_0)$ can be expressed  as the limit for $dt\rightarrow 0$ of the minimum value that the finite entropy increments $[S(\Phi_{dt}(\rho)) -S(\Phi_{0}(\rho))]/dt= [S(\Phi_{dt}(\rho)) -S(\rho)]/dt$ assume over the
set $\mathfrak{S}_{S_0}({\cal H}_A)$. Of course if {\bf (v2)} is true then for all $dt$ such minimum is achieved on $\rho_0$, validating Eq.~(\ref{infinitesimal}).
Analogously if the infinitesimal version {\bf (v3)}  is true for all
$S_0$, then all the  channels $\Phi_T$ generated by ${\cal L}$ will
obey the finite version {\bf (v2)}  of the conjecture (to verify
this simply use the fact that Gibbs states are mapped into Gibbs
states).

Proving or disproving any of the  above version of the conjecture is
apparently a formidable task. In the following section we will then
focus on some special (non trivial) cases for which some preliminary
results can be  derived.

\section{A special case}\label{s:special}
As a special class of single mode Gaussian  channel $\Phi$ consider
an attenuator channel ${\cal E}_\eta^N$ which mixes via a
beam-splitter (BS) of transmissivity $\eta \in [0,1]$ the input
state of the system with a (thermal) Gibbs environmental state
$\rho_E$ characterized by having $N$ mean photon number and thus
entropy
\begin{eqnarray} \label{entropyE}
S(\rho_E) =g(N) \;,
\end{eqnarray}
with $g(\cdot)$ as in Eq.~(\ref{entropyN}).
 In the language of Ref.~\cite{CANONICAL} this map is attenuator belonging to the class C
 with
$\kappa^2 =\eta$ and $c= (1-\eta) N$ in Eq.~(\ref{dfe}). Alternatively, following the notation of
 Ref.~\cite{CONJEC}, it can be expressed in terms of the following
input-output transformation
\begin{eqnarray} \label{neweq1}
\chi(\mu)  \longrightarrow \chi'(\mu) = \chi(\sqrt{\eta} \mu) \; e^{ - (1-\eta) (N+1/2) |\mu|^2 }\;,
\end{eqnarray}
where   $\chi(\mu) = \mbox{Tr} [ \rho D(\mu)]$  and  $  \chi'(\mu)
=\mbox{Tr} [ {\cal E}_\eta^{N}(\rho) D(\mu)]$ are the symmetrically
ordered characteristic function of the input and output state of the
system, respectively ($D(\mu) =\exp[ \mu a^\dag - \mu^* a]$ being
the displacement operator of the mode). This channel maps the Gibbs
state $\rho_0$ into a new Gibbs state $\rho'_0$ of average photon
number $N'_0=\eta N_0 + (1-\eta)N$ and of output entropy
\begin{eqnarray}
S({\cal E}_\eta^{N}(\rho_0)) = S(\rho_0')= g(N'_0)= g(\eta N_0 + (1-\eta)N).
\end{eqnarray}
According to the version {\bf (v2)} of the  conjecture then we
should have ${\cal F}({\cal E}_\eta^{N}; S_0) = g(\eta N_0 +
(1-\eta)N)$, or equivalently
\begin{eqnarray}
S({\cal E}_\eta^{N}(\rho)) \geqslant   g(\eta N_0 + (1-\eta)N) \;,\label{ines}
\end{eqnarray}
for all $\rho\in  \mathfrak{S}_{S_0}({\cal H})$.
Also proving this inequality is rather complicated. In the following we thus focus on the following  (very) specific configuration where
the input entropy $S_0$ which defines the set
$\mathfrak{S}_{S_0}({\cal H})$ of possible input states  coincides
with the entropy of the environment $\rho_E$. In particular due to
Eqs.~(\ref{entropyN}) ad (\ref{entropyE})  this implies that
$\rho_E$ and $\rho_0$ are indeed the same state, and thus
\begin{eqnarray}
g(N_0) = g(N) \Longleftrightarrow N= N_0\;.
\end{eqnarray}
Under this condition we first notice that $\rho_0$ is the fixed point of the map ${\cal E}_\eta^{N_0}$, i.e.
\begin{eqnarray}
{\cal E}_\eta^{N_0}(\rho_0) = \rho_0\;,
\end{eqnarray}
 (this can be easily verified from Eq.~(\ref{neweq1}) by reminding that the symmetrically ordered characteristic function of the Gibbs state $\rho_0$ is
 $\exp[ -  (N_0+1/2) |\mu|^2 ]$). 
Therefore proving the inequality~(\ref{ines}) (and thus the conjecture {\bf v(2)}) is now equivalent to showing that
\begin{eqnarray} \label{ffd}
S({\cal E}_\eta^{N_0}(\rho)) \geqslant   g(N_0) \;,
\end{eqnarray}
holds for all  $\rho\in  \mathfrak{S}_{S_0}({\cal H})$.
Equivalently, this can also be rewritten as  (see Eq.~(\ref{dds})),
\begin{eqnarray}
\eta  S(\rho||\rho_0) \geqslant  S({\cal E}_\eta^{N_0}(\rho)||{\cal E}_\eta^{N_0}(\rho_0))\;,\label{ddssdd}
\end{eqnarray}
which should again apply to all  $\rho\in  \mathfrak{S}_{S_0}({\cal
H})$. Among the various properties of the channel ${\cal
E}_\eta^{N_0}$ we remind that they form a semigroup under
multiplication due to the properties~\cite{CANONICAL1,CONJEC}
\begin{eqnarray}\label{semi}
{\cal E}_{\eta_1}^{N_0} \circ {\cal E}_{\eta_2}^{N_0}  = {\cal E}_{\eta_2 \eta_1}^{N_0}\;,   \qquad \qquad {\cal E}_{\eta=1}^{N_0} ={\cal I}\;.
\end{eqnarray}
Defining thus $\Phi_t = {\cal E}_{\eta = e^{-t}}^{N_0}$ the Lindblad
generator is easily derived as in Eq.~(\ref{ffd111}) with $\gamma_+
= N_0$ and $\gamma_- = N_0+1$. This allows us to rephrase the
infinitesimal version {\bf (v3)} of the conjecture as
\begin{eqnarray}
{\cal F}({\cal L}; S_0) = - \mbox{Tr}[ {\cal L} (\rho_0) \ln \rho_0  ] = 0
 \;. \label{f1}
\end{eqnarray}
Interestingly enough even though for arbitrary values of $\eta$ the inequality~(\ref{ffd}) is
 difficult to derive, there are some special case in which it simply follows by
general consideration on von Neumann entropy. Specifically the following result can be shown:
\newline \\
{\bf Theorem.} {\em For arbitrary positive values of $N_0\geqslant
0$, the inequalities in {\em Eqs.~(\ref{ffd})} and  {\em (\ref{ddssdd})}
hold for all $\eta = 1/k$ with $k$ integer.}
\newline \\
{\em Proof:} For $k=1$ the result is trivial. For $k\geqslant 2$ it
follows from the subadditivity of the von Neumann entropy. In
particular consider first the case of $k=2$. In this case we
introduce a unitary representation~\cite{CANONICAL1,CANONICAL}  of
the channel ${\cal E}_{1/2}^{N_0}$ constructed by mixing the input
state $\rho$ via a BS of transmissivity $\eta=1/2$ with the thermal
environment $\rho_E$, i.e.
\begin{eqnarray}
{\cal E}_{1/2}^{N_0}(\rho) = \mbox{Tr}_E [ U_{1/2}^{(AE)} (\rho \otimes \rho_E) [U_{1/2}^{(AE)}]^\dag ] \;, \label{fd1e}
\end{eqnarray}
where  $\mbox{Tr}_E$ is the partial trace over the environment and where
$U_{\eta}^{(AE)} = \exp[ \arccos\sqrt{\eta} (a^\dag b - a b^\dag)]$ is the BS unitary coupling which connects $A$ and $E$ (here $a$ and $b$ stands for the annihilation operators of the two systems).
 In this case the weak complementary $\tilde{\cal E}_{1/2}^{N_0}(\cdots)= \mbox{Tr}_A [ U_{1/2}^{(AE)} (\cdots \otimes \rho_E) [U_{1/2}^{(AE)}]^\dag ]$  
   is known~\cite{fazio,CANONICAL1}  to be unitary equivalent to ${\cal E}_{1/2}^{N_0}$ (here $\mbox{Tr}_A$ indicates the partial trace over the system degree of freedom). Therefore
by invoking the subadditivity of the von Neumann entropy we can
write,
\begin{eqnarray}
&2 S({\cal E}_{1/2}^{N_0}(\rho)) = S({\cal E}_{1/2}^{N_0}(\rho)) + S(\tilde{\cal E}_{1/2}^{N_0}(\rho))& \nonumber  \\ \nonumber
\geqslant
& S(U_{1/2}^{(AE)} (\rho \otimes \rho_E) [U_{1/2}^{(AE)}]^\dag) = S(\rho) + S(\rho_E) = 2 g(N_0),&
\end{eqnarray}
which proves the thesis (in the last identity we used the fact that
since  $\rho\in   \mathfrak{S}_{S_0}({\cal H})$ it has the same
entropy $S_0=g(N_0)$ of  $\rho_E$). For $k>2$  we use a similar
trick concatenating more BS transformations in series in order to
obtain a set-up with $k$ output ports and $k$ inputs (one input for
the state $\rho$ and  the remaining  for $k-1$ copies of $\rho_E$).
Adjusting the transmissivities of the BS in such a way to guarantee
that all of output ports  have  overall transmissivities $1/k$ we
can invoke the subadditivity to finally derive the inequality
\begin{eqnarray} S({\cal E}_{1/k}^{N_0}(\rho)) \geqslant g(N_0)\;,
\label{res}
\end{eqnarray}
which proves the thesis. More precisely the above construction consists in  introducing $k-1$ copies of the state $\rho_E$ and introducing  the following $k$ modes state,
\begin{eqnarray}
\Omega_{AE_1\cdots E_{k-1}} = \rho \otimes \rho_{E_1} \otimes \rho_{E_2} \otimes \cdots \otimes \rho_{E_{k-1}} \;,
\end{eqnarray}
which has entropy equal to $k g(N_0)$ when $\rho\in \mathfrak{S}_{S_0}({\cal H}_A)$.
Consider then following unitary couplings
\begin{eqnarray}
W = U_{\eta_{k-1}}^{AE_{k-1}} \cdots U_{\eta_2}^{AE_{2}}  U_{\eta_1}^{AE_{1}} \;,
\end{eqnarray}
where for $j=1, \cdots, k-1$,  $U_{\eta_{j}}^{AE_{j}}$ is the BS unitary transformation of
transmissivity $\eta_{j}$ which couples $A$ with the  system $E_j$. 
The inequality (\ref{res}) then can be obtained applying the
subadditivity of von Neumann entropy to the state
$\Omega'_{AE_1\cdots E_k}= W \Omega_{AE_1\cdots E_k} W^\dag$, i.e.
using the relation
\begin{eqnarray}\label{master}
S(\Omega'_{AE_1\cdots E_{k-1}}) \leqslant S(\Omega'_A)+  \sum_{j=1}^{k-1} S(\Omega'_{E_{j}}) \;,
\end{eqnarray}
where $\Omega'_A$ is the reduced matrix of $\Omega'_{AE_1\cdots E_k}$ associated with the system $A$, and where for all $j=\{ 1, \cdots, k-1\}$
$\Omega'_{E_{j}}$ is the reduced matrix of  $\Omega'_{AE_1\cdots E_k}$ associated with the system $E_j$.
Indeed the left-hand side term of this expression coincides with the von Neumann entropy of $\Omega_{AE_1\cdots E_k}$ hence
\begin{eqnarray}
S(\Omega'_{AE_1\cdots E_{k-1}}) = k g(N_0)\;.
\end{eqnarray}
On the other hand we notice that for reduced density operator of the subsystem $A$ one has,
\begin{eqnarray}
\Omega'_A &\equiv& \mbox{Tr}_{E_1\cdots E_{k-1}} [ \Omega'_{AE_1\cdots E_{k-1}} ] \nonumber \\
&=& {\cal E}_{\eta_{k-1}}^{N_0} \circ {\cal E}_{\eta_{k-2}}^{N_0} \circ \cdots \circ {\cal E}_{\eta_1}^{N_0} (\rho) = {\cal E}_{\bar{\eta}_{k}}^{N_0}(\rho)\;,
\end{eqnarray}
where we used the semigroup property (\ref{semi}) and defined $\bar{\eta}_{k}\equiv \eta_{k-1} \eta_{k-2} \cdots \eta_2\eta_1$.
Similarly for the reduced density operator associated with the system $E_1'$ we notice that
\begin{eqnarray}
\Omega'_{E_1} &\equiv& \mbox{Tr}_{AE_2 E_3\cdots E_{k-1}} [ \Omega'_{AE_1\cdots E_{k-1}} ] \nonumber \\
&=&  \mbox{Tr}_{A} [ U_{\eta_1}^{(AE_1)} (\rho\otimes \rho_{E_1})  [U_{\eta_1}^{(AE_1)}]^\dag  ]  = \tilde{\cal E}_{\eta_1}^{N_0}(\rho)\;, \nonumber
\end{eqnarray}
with $\tilde{\cal E}_{\eta_1}^{N_0}$ being the weak-complementary of
the channel ${\cal E}_{\eta_1}^{N_0}$ under the unitary
representation of Eq.~(\ref{fd1e}). Apart from an irrelevant unitary
rotation, this is know to be equivalent to the channel ${\cal
E}_{1-\eta_1}^{N_0}$~\cite{fazio,CANONICAL1}. Thus we can conclude that,
\begin{eqnarray}
S(\Omega'_{E_1}) = S( \tilde{\cal E}_{\eta_1}^{N_0}(\rho)) = S({\cal E}_{1-\eta_1}^{N_0}(\rho)) \;.
\end{eqnarray}
In a similar fashion we have that for arbitrary $j=1,2, \cdots, k-1$ the reduced density matrices of the subsystem $E_j$ can be expressed as
\begin{eqnarray}
\Omega'_{E_j} &=&  \tilde{\cal E}_{\eta_j}^{N_0}\circ {\cal E}_{\eta_{j-1}}^{N_0}\circ {\cal E}_{\eta_{j-2}}^{N_0}\circ\cdots \circ {\cal E}_{\eta_{1}}^{N_0}(\rho)\nonumber \\
&=& \tilde{\cal E}_{\eta_j}^{N_0}\circ {\cal E}_{\eta_{j-1}\eta_{j-2} \cdots \eta_1}^{N_0}(\rho)
\;, \nonumber  \end{eqnarray}
where again the semigroup property (\ref{semi}) was used to simplify the expression. Exploiting then the unitary equivalence between  $\tilde{\cal E}_{\eta_j}^{N_0}$ and  ${\cal E}_{1-\eta_j}^{N_0}$ we finally get
\begin{eqnarray}
S(\Omega'_{E_j}) &=&S(\tilde{\cal E}_{\eta_j}^{N_0}\circ {\cal E}_{\eta_{j-1}\eta_{j-2} \cdots \eta_1}^{N_0}(\rho))\\\nonumber
& =& S({\cal E}_{1-\eta_j}^{N_0}\circ {\cal E}_{\eta_{j-1}\eta_{j-2} \cdots \eta_1}^{N_0}(\rho))
=S({\cal E}_{\bar{\eta}_j}^{N_0}(\rho))
\;,\end{eqnarray}
with $\bar{\eta}_j = (1-\eta_j)\eta_{j-1} \eta_{j-2} \cdots \eta_{1}$. Equation~(\ref{master}) can thus be rewritten as,
\begin{eqnarray}\label{master1}
k g(N_0)  \leqslant \sum_{j=1}^{k}  S({\cal E}_{\bar{\eta}_j}^{N_0}(\rho))\;.
\end{eqnarray}
To prove the thesis we need thus only to find $\eta_j$ such that $\bar{\eta}_j=1/k$ for all $j = 1, 2, \cdots, k$.
To do so we take  $\eta_j= \frac{k-j}{k-j+1}$ for all $j={1,2, \cdots, k-1}$.
With this choice the right-hand side term of
 Eq.~(\ref{master1}) becomes $\sum_{j=1}^{k}  S({\cal E}_{\bar{\eta}_j}^{N_0}(\rho)) = k S({\cal E}_{1/k}^{N_0}(\rho))$
 yielding Eq.~(\ref{res}). $\blacksquare$

 \section{Proof  of the conjecture for one mode degenerate Gaussian channels}\label{singular}

In this section we describe the solution of the conjecture {\bf (v1)} in the cases of one mode degenerate Gaussian  channels. In the canonical form
of Ref.~\cite{CANONICAL,CANONICAL1} they correspond to the classes $A_1, A_2, B_1$ and are formally characterized by the fact
 at least one of the two $2\times 2$ matrices that describe their action on the Weyl operator of the system is not
invertible.

Channels belonging to the class
$A_1$ satisfy the equation
\begin{equation}
\chi(\mu)  \longrightarrow \chi'(\mu) = \chi(0) \; e^{ - (N+1/2) |\mu|^2 }\;,
\end{equation}
which maps any input state into fixed output Gaussian state (indeed
they can be seen are limiting cases of  attenuators channels (class
$C$) with zero beam splitter transmissivity). Hence for these
channels, the output entropy is constant and the problem is trivial.

Case $A_2$ corresponds to the equation 
\begin{equation}\label{chiprime}
\chi(\mu)  \longrightarrow \chi'(\mu) = \chi(-i\Im \mu) \; e^{ -
(N+1/2) |\mu|^2 }\;,
\end{equation}
where $\Im \mu$ is the imaginary part of $\mu$. The channel is given
explicitly by 
\begin{equation}\label{A2}
\Phi (\rho )=\int e^{ ixp}\rho_Ee^{ - ixp}\; P_{\rho}(dx),
\end{equation}
  where $\rho_E$ is a Gibbs state of mean energy $N$,   $p= i (a^\dag - a)/\sqrt{2}$ is the momentum quadrature of the system, and 
 $P_{\rho}(dx)=\langle x|\rho |x\rangle dx$ is the probability
distribution of the position operator $q=(a^\dag + a)/\sqrt{2}$ in the state
$\rho$~\cite{nota1}.
It is  an entanglement-breaking channel~\cite{ENTB} which describes position measurement followed by preparation of the state $e^{ ixp}\rho_Ee^{- ixp}$ shifted by the outcome of the measurement $x$. By concavity of the entropy
\begin{equation}
S(\Phi (\rho ))\geqslant \int S(e^{ixp}\rho_Ee^{ - ixp})P_{\rho}(dx)=S(\rho_E),
\end{equation}
and in fact
\begin{equation}
\inf_{\rho \in \mathfrak{S}_{S_0}({\cal H}_A)}S(\Phi (\rho ))=S(\rho_E).
\end{equation}
To prove this, consider the input Gaussian states $\rho_{\sigma_q,\sigma_p}$ with zero mean, variances $\mathsf{D}q=\sigma^2_q, \mathsf{D}p=\sigma^2_p,$ and uncorrelated $q,p$.
The entropy of such states is equal to $S(\rho_{\sigma_q,\sigma_p})=g\left(\frac{\sigma_q\sigma_p}{\hbar}-\frac{1}{2}\right)$. By fixing it equal to $S_0$ and letting $\sigma_q\to 0$, we obtain $S(\rho')\to S(\rho_E)$.

Case $B_1$ is described by the equation 
\begin{equation}
\chi(\mu)  \longrightarrow \chi'(\mu) = \chi(\mu) \; e^{ -
(1/2)|\Im\mu|^2 }\;,
\end{equation}
which corresponds to degenerate additive Gaussian classical noise (only in the component $q$, with variance $1/2$). In other words
\begin{equation}
\Phi (\rho )=\int e^{ ixp}\; \rho \; e^{ - ixp}P (dx),
\end{equation}
 where $P (dx) =dx  \exp[- x^2/4]/\sqrt{4\pi} $ is a Gaussian noise distribution. Then similarly to the previous case, $S(\rho')\geqslant S(\rho)=S_0$. Moreover
\begin{equation}
\Phi (\rho_{\sigma_q,\sigma_p} )= \rho_{\sqrt{\sigma^2_q+1/2},\sigma_p}
\end{equation}
so fixing $S(\rho_{\sigma_q,\sigma_p})=S_0$ and letting $\sigma_p\to 0$, we obtain $S(\rho')\to S_0$.
Thus
\begin{equation}
\inf_{\rho \in \mathfrak{S}_{S_0}({\cal H}_A)}S(\Phi (\rho ))=S_0.
\end{equation}

\section{Conclusion}\label{s:conc}
In this work we discussed a generalized minimal output
conjecture for Gaussian channels. For degenerate one-mode quantum channels it has
been proved explicitly. For attenuator channels the conjecture was
proved for some values of the transmissivity,  under the assumption
that the input entropy and the entropy of the thermal state
environment coincide.

VG acknowledges support from the FIRB-IDEAS project under the
contract RBID08B3FM. AH acknowledges the support of Scuola Normale
Superiore di Pisa, which he was visiting while part of this work was
done, and the RAS Program ``Mathematical Control Theory''. SL acknowledges the support
NEC, DARPA, ONR, and the Keck Foundation.


\begin{thebibliography}{1}
\bibitem{nota1}  To verify that the operator (\ref{A2}) has the
characteristic function $\chi'(\mu)$ in the right side of
(\ref{chiprime}) use the consequence of the Canonical Commutation
Relation $\exp(-ixp)D(\mu)\exp(ixp)=\exp(-ix\sqrt{2}\Im \mu)D(\mu) $
and the fact that  $\int\exp(-ix\sqrt{2}\Im \mu) |x\rangle \langle x| dx=D(-i\Im\mu).$
\end{thebibliography}
\end{document}